\documentclass[a4paper,11pt]{revtex4}
\usepackage{graphicx}
\usepackage{amsmath}
\usepackage{hyperref}
\usepackage{subfigure}
\usepackage{amssymb}
\usepackage{dcolumn}
\usepackage{bm}

\begin{document}

\title{Nonequilibrium effects in dynamic symmetry breaking}

\author{Marlene Nahrgang}
\affiliation{SUBATECH, UMR 6457, Universit\'e de Nantes, Ecole des Mines de Nantes,
IN2P3/CRNS. 4 rue Alfred Kastler, 44307 Nantes cedex 3, France}
\affiliation{Frankfurt Institute for Advanced Studies (FIAS), Ruth-Moufang-Str.~1, 60438 Frankfurt am Main, Germany}
\author{Stefan Leupold}
\affiliation{Department of Physics and Astronomy, Uppsala University, Box 516, 75120 Uppsala, Sweden}
\author{Marcus Bleicher}
\affiliation{Frankfurt Institute for Advanced Studies (FIAS), Ruth-Moufang-Str.~1, 60438 Frankfurt am Main, Germany}
\affiliation{Institut f\"ur Theoretische Physik, Goethe-Universit\"at, Max-von-Laue-Str.~1,
 60438 Frankfurt am Main, Germany}

\begin{abstract}
We present a dynamic model based on the linear sigma model with constituent quarks that allows for studying the explicit propagation of fluctuations at a chiral critical point and a first-order phase transition.
The coupled dynamics of the sigma field and the quarks is derived selfconsistently. Hereby, the sigma field evolves according to a semiclassical Langevin equation of motion, while the evolution of the quarks is described fluid dynamically in order to model the expansion of a hot fireball in heavy-ion collisions.
Dissipative processes and fluctuations in the Langevin equation are allowed under the assumption of energy-momentum conservation of the coupled system.
We study the evolution of the constituent quark masses in this nonequilibrium set up.
\end{abstract}

\maketitle

\section{Introduction}
The QCD phase diagram is expected to exhibit a rich phase structure. At larger baryochemical potential, as achieved at the upcoming FAIR project \cite{Friman:2011zz}
 at GSI Darmstadt, a first order phase transition is expected from model studies \cite{Scavenius:2000qd,Ratti:2005jh,arXiv:0704.3234}. Interesting observables could here be based on the growth of fluctuations due to the nonequilibrium effect of supercooling leading to nucleation and spinodal decomposition \cite{Csernai:1992tj,Mishustin:1998eq,Randrup:2010ax,Chomaz:2003dz}.

At zero baryochemical potential the nature of the phase transition of QCD is well understood from lattice QCD calculations, which show that it is an analytic crossover \cite{Aoki:2006we}. As a consequence there must be a critical point, which terminates the line of first order phase transitions. Unfortunately, there are no analytic techniques to solve the QCD partition function in this regime. Thus, one has to rely on model calculations and explore the phase diagram in experiment. In equilibrium systems fluctuations and correlations of the order parameter diverge at the critical point. Coupling particles to the sigma field, the order parameter of chiral symmetry, leads to a nonmonotonic behaviour in fluctuations of net-charge or net-baryon number multiplicities  \cite{Stephanov:1998dy,Stephanov:1999zu}. The key ingredient is the correlation length which becomes infinite in a system at a critical point. In a realistic evolution of a heavy-ion collision, however, the growth of the correlation length is limited by the size of the system and by the finite time, which the dynamic systems spends at a critical point. Relaxation times also become infinite at the critical point, a phenomenon called critical slowing down. Even if the system is in equilibrium above the critical point it is necessarily driven out of equilibrium by passing trough the critical point. Assuming a phenomenological time evolution of the correlation length with parameters from the $3$d Ising universality class it was found that the correlation length does not grow beyond $2-3$ fm \cite{Berdnikov:1999ph}.

The explicit propagation of fluctuations coupled to a dynamic model is a necessary step towards understanding the QCD phase diagram from heavy-ion collision experiments. 
In chiral fluid dynamic models \cite{Mishustin:1998yc,Paech:2003fe} the propagation of the fields in the chiral sector is coupled to a fluid dynamic propagation of the constituent quarks. The expansion and cooling of the fluid does thus drive the underlying model through the phase transition.
In the following we present the model of nonequilibrium chiral fluid dynamics \cite{Nahrgang:2011mg} with a focus o the evolution of the constituent quark masses.

\section{Nonequilibrium chiral fluid dynamics}

\begin{figure}
\includegraphics[width=0.5\textwidth]{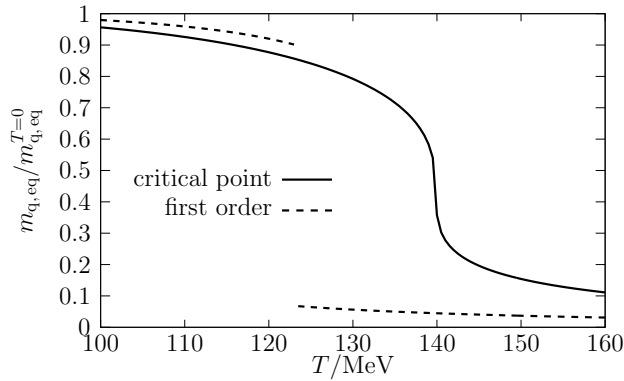}
\label{fig:mqeq}
\caption{The equilibrium quark masses as a function of temperature for a critical point and a first order phase transition}
\end{figure}

Starting from the quark meson model \cite{Schaefer:2004en} the coupled dynamics of the order parameter of chiral symmetry, the sigma field, and the fluid dynamic expansion of the quarks have been derived \cite{Nahrgang:2011mg}.
The Langevin equation for the sigma mean-field reads
\begin{equation}
 \partial_\mu\partial^\mu\sigma+\frac{\delta V_{\rm eff}}{\delta\sigma}+\eta\partial_t \sigma=\xi\, .
\label{eq:equi_langevineq}
\end{equation}
The effective potential to one-loop level is given by
\begin{equation} 
V_{{\rm eff}}(\sigma,T)=U\left(\sigma \right)-2d_q T \int\frac{{\rm d}^3p}{(2\pi)^3}\ln\left(1+\exp\left(-\frac{E}{T}\right)\right) \, ,
\label{eq:effpot}
\end{equation}
The classical potential $U$ in (\ref{eq:effpot}) is of Mexian hat shape
\begin{equation}
U\left(\sigma \right)=\frac{\lambda^2}{4}\left(\sigma^2-\nu^2\right)^2-h_q\sigma-U_0\, .
\label{eq:Uchi}
\end{equation} 
The parameters are chosen such that chiral symmetry is spontaneously broken in the vacuum, where $\langle\sigma\rangle=f_\pi=93$~MeV. Chiral symmetry is broken explicitly by the term $h_q\sigma$. In the effective potential $V_{\rm eff}$ the quark energy is given by 
\begin{equation}
E^2=p^2+m_q^2
\end{equation}
and the quark mass is generated dynamically at the phase transition, where the sigma develops a finite expectation value and thus $m_q=g\langle\sigma\rangle$. As due to the explicit symmetry breaking term $\langle\sigma\rangle$ does not vanish exactly in the chirally restored phase, but has a small finite value, $m_q$ does not exactly vanish either. The effective potential is given for $\mu_B=0$, where the correct vacuum value for the constituent quark mass is obtained by a coupling of $g=3.3$ between the sigma field and the quark fields. For a qualitative study we fix the baryochemical potential at $\mu_B=0$ and tune the phase transition by changing the coupling constant. For $g=3.63$ one finds a corresponding critical point at $T_c=139.88$ MeV by the vanishing curvature of the effective potential at the minimum and a first order phase transition for $g=5.5$ and $T_c=123.27$ MeV by the appearance of two degenerate minima. 
For these couplings the phenomenologically known value of the constituent quark mass comes out wrong. We, however, focus on the qualitative behaviour and leave the extension of the model to finite $\mu_B$ to future work. In figure \ref{fig:mqeq} we, therefore, show the equilibrium quark masses devided by the vacuum quark masses. For a first order phase transition we see a large discontinuity at the phase transition due to the two degenerate minima, while the transition is smooth at the critical point.
The damping term is \cite{Greiner:1996dx,Biro:1997va,Rischke:1998qy,Nahrgang:2011mg}
\begin{equation}
 \eta=\begin{cases}
        g^2\frac{d_q}{\pi}\left(1-2n_{\rm F}\left(\frac{m_\sigma}{2}\right)\right)\frac{1}{m_\sigma^2}\left(\frac{m_\sigma^2}{4}-m_q^2\right)^{3/2}  & \text{for }m_\sigma>2m_q=2g\sigma_{\rm eq}\\
2.2/{\rm fm}  & \text{for }2m_q>m_\sigma>2m_\pi\\
0 & \text{for }m_\sigma<2m_\pi,2m_q
      \end{cases}
\end{equation}
The stochastic field in the Langevin equation (\ref{eq:equi_langevineq}) has a vanishing expectation value
\begin{equation}
 \langle\xi(t)\rangle_\xi=0\, ,
\end{equation}
and the noise correlation is given by the dissipation-fluctuation theorem
\begin{equation}
 \langle\xi(t)\xi(t')\rangle_\xi=\frac{1}{V}\delta(t-t')m_\sigma\eta\coth\left(\frac{m_\sigma}{2T}\right)\, .
\label{eq:sc_noisecorrelation}
\end{equation}
The local pressure and energy density of the quarks is given by 
\begin{equation}
  p(\sigma,T)= -V_{\rm eff}(\sigma,T)+U(\sigma)\quad\quad e(\sigma,T)= T\frac{\partial p(\sigma,T)}{\partial T}-p(\sigma,T)\; .
\label{eq:lsm_eos1}
\end{equation}
In the relativistic fluid dynamic equations we find a source term $S^\nu$ allowing for the energy dissipation from the system to the heat bath
\begin{equation}
\partial_\mu T^{\mu\nu}=S^\nu\, .
\label{eq:fluidT}
\end{equation}

\section{Dynamic quark masses}
We solve equations (\ref{eq:equi_langevineq}) and (\ref{eq:fluidT}) numerically for simple initial conditions: an almond-shape, boost-invariant distribution of the energy density in the transverse plane and longitudinal direction respectively. First numerical results have been presented in \cite{arXiv:1105.1962}.
\begin{figure}
\subfigure{\includegraphics[width=0.45\textwidth]{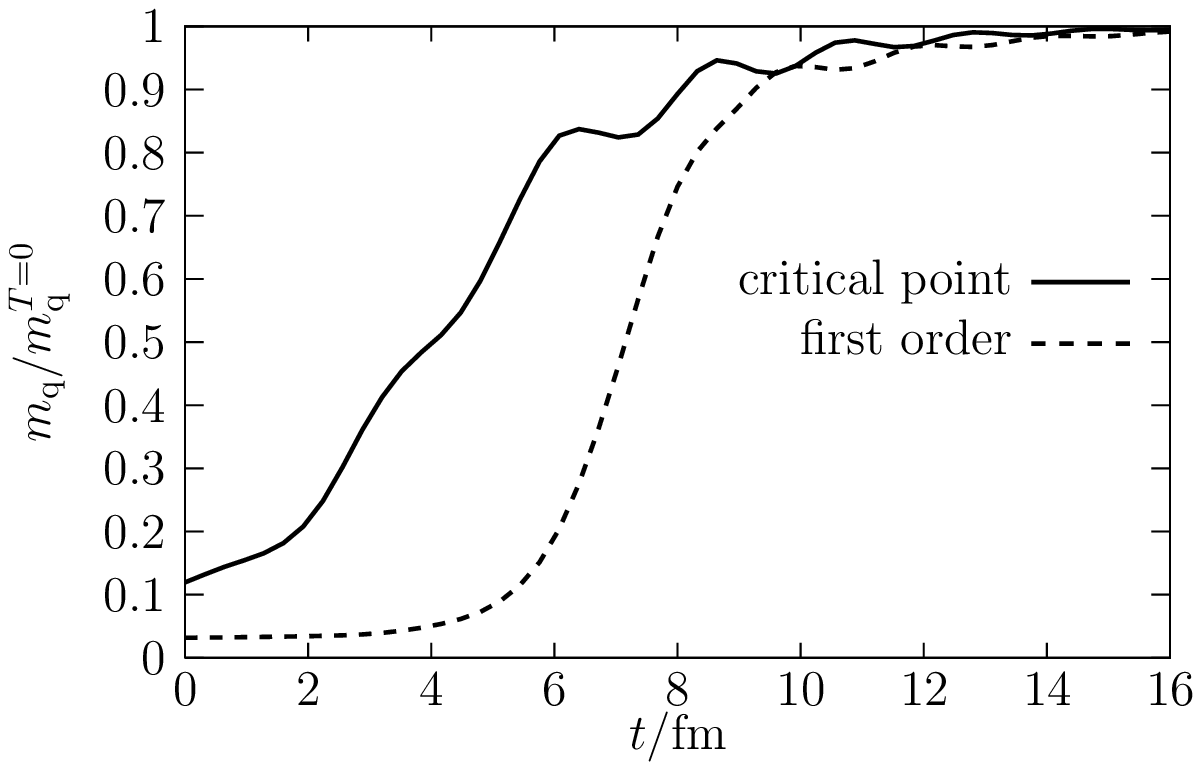}}
\subfigure{\includegraphics[width=0.45\textwidth]{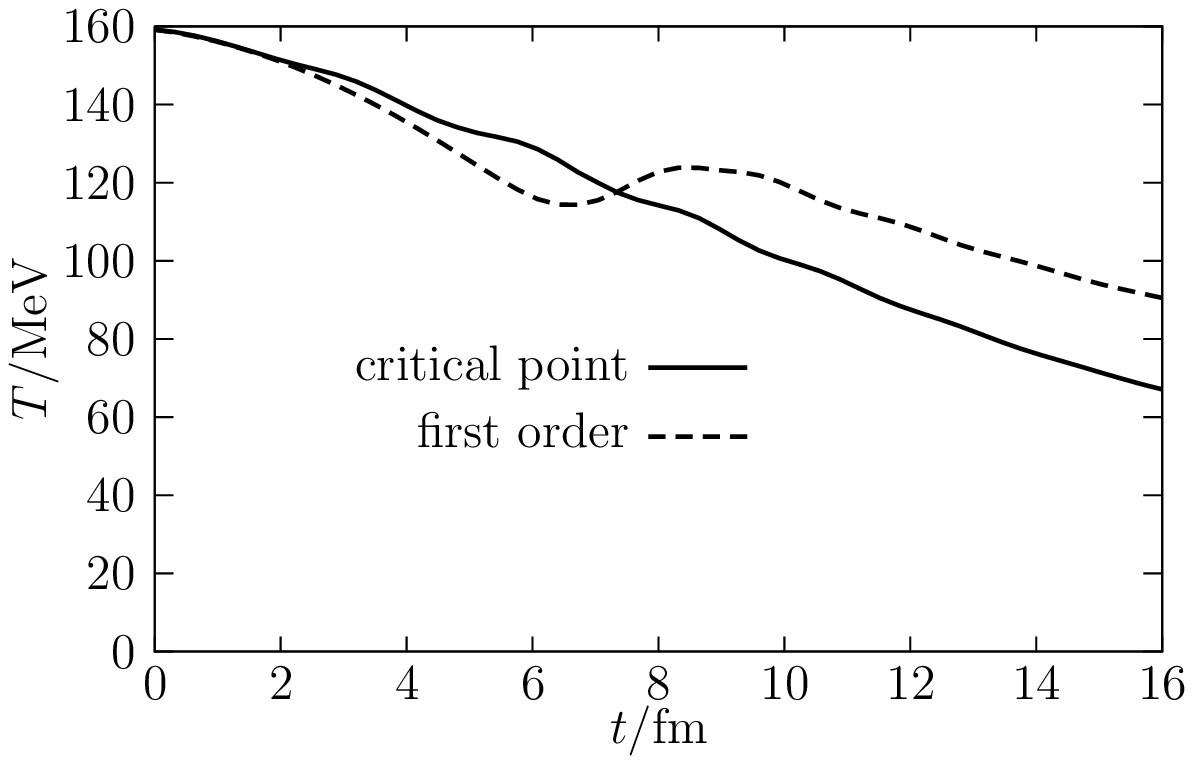}}
\label{fig:mqnoneq}
\caption{Evolution of the constituent quark masses for a scenario with a critical point and a first order phase transition (left). The temperature evolution is shown for comparison (right).}
\end{figure}
In figure \ref{fig:mqnoneq} (left) we present the time evolution of the constituent quarks mass during the expansion of the quark fluid and the relaxation of the sigma field for both scenarios, with a critical point and with a first order phase transition.
For a better understanding we also show the average temperature of the quark fluid in the same rregion. The transition temperature is crossed around $t\approx 5$~fm for a first order phase transition and around $t\approx 4$~fm for a critical point. At the first order phase transition we observe the interesting effect of reheating between $5$ and $10$~fm. In \cite{Nahrgang:2011ll} we discussed how this effect is caused due to the explicit energy-momentum conservation of the coupled system.
We see that during the evolution of the system the constituent quark masses increase to their vacuum value. For a critical point scenario this behaviour is rather smooth and begins even at temperatures above the phase transition as the effective potential changes its shape gradually. During a first order phase transition, however, the high-temperature expectation value is separated from the vacuum expectation value by a finite barrier. Parts of the system are trapped in the unstable minimum below the phase transition. The jump of the constituent quark masses at the first order phase transition are thus smirred out and deviations from the equilibrium form are larger than in a critical point scenario.

\section{Summary}
We have presented a model of chiral fluid dynamics including dissipation and noise based on the linear sigma model with constituent quarks. The latter receive their vacuum mass by a finite expectation value of the sigma field in the chirally broken phase. We and studied the time evolution of the constituent quark mass. Due to supercooling at the first order phase transition we observe a retarded approach to the vacuum value. There are substantial parts below the phase transition where the quarks in the fluid have small masses compared to $m_q^{T=0}$. It will be interesting to study the influence of this effect in coupling to the confinement-deconfinement phase transition. Work including the Polyakov-loop is in progress \cite{inprogress}.

This work was supported by the Hessian Excellence Inititive LOEWE through the Helmholtz International Center for FAIR.

\end{document}